\documentclass[10pt,conference]{IEEEtran}
\usepackage[centertags]{amsmath}
\usepackage{amsfonts}
\usepackage{amssymb}
\usepackage{epsfig}
\usepackage{amsmath,  amssymb}
\usepackage{graphicx}
\usepackage[centertags]{amsmath}
\usepackage{clrscode}
\usepackage{cite}
\usepackage{fullpage}
\usepackage{amsmath,amssymb,mathrsfs,pseudocode}
\usepackage{color}
\usepackage[normalem]{ulem}
\usepackage{psfrag}

\usepackage[margin=18.5mm]{geometry}
\usepackage{xspace}
\usepackage{multicol}
\usepackage{booktabs} 
\usepackage{multirow}
\usepackage[x11names,dvipsnames,table]{xcolor}

\newcommand{\myhline}{\hline}
\newcommand{\opertype}[1]{\begin{minipage}{29mm}\centering\vspace{1mm} #1\vspace{1mm}\end{minipage}}

\newtheorem{theorem}{Theorem}

\newtheorem{example}[theorem]{Example}
\newtheorem{corollary}[theorem]{Corollary}

\usepackage{xspace}
\usepackage{bbm}
\input{mathlig}

\newcommand{\muspace}{\mspace{1mu}}

\DeclareRobustCommand{\scond}{\mathchoice{\muspace\vert\muspace}{\vert}{\vert}{\vert}}
\mathlig{|}{\scond}
\newcommand{\cond}{\,\vert\,}

\DeclareRobustCommand{\discint}{\mathchoice{\mspace{-1.5mu}:\mspace{-1.5mu}}{\mspace{-1.5mu}:\mspace{-1.5mu}}{:}{:}}
\mathlig{::}{\discint}
\newcommand{\suchthat}{\colon}

\newcommand{\Cr}{\mathscr{C}}

\newcommand{\Cv}{{\bf C}}

\newcommand{\Rv}{{\bf R}}

\newcommand{\xh}{{\hat{x}}}

\DeclareMathOperator\R{\textsf{R}}

\def\textiid{i.i.d.\@\xspace}
\newcommand\iid{\ifmmode\text{ i.i.d. } \else \textiid \fi}

\def\clap#1{\hbox to 0pt{\hss#1\hss}}
\def\mathclap{\mathpalette\mathclapinternal}
\def\mathclapinternal#1#2{%
  \clap{$\mathsurround=0pt#1{#2}$}}

\let\oldstackrel\stackrel
\renewcommand{\stackrel}[2]{\oldstackrel{\mathclap{#1}}{#2}}

\begin{document}
\title{Distributed Index Coding\vspace{-2mm}}

\author{Parastoo Sadeghi$^{\dag}$, Fatemeh Arbabjolfaei$^{*}$, and Young-Han Kim$^{*}$\\\vspace{-0mm}
$^{\dag}$Research School of Engineering, Australian National University, Canberra, ACT, 2601, Australia\\
$^{*}$Department of Electrical and Computer Engineering, University of California, San Diego, CA 92093, USA\\
Emails:  parastoo.sadeghi@anu.edu.au, \{farbabjo, yhk\}@ucsd.edu}


\maketitle

\begin{abstract}
In this paper, we study the capacity region of the general distributed index coding. In contrast to the traditional centralized index coding where a single server contains all $n$ messages requested by the receivers, in the distributed index coding there are $2^n-1$ servers, each containing a unique non-empty subset $J$ of the messages and each is connected to all receivers via a noiseless independent broadcast link with an arbitrary capacity $C_J \ge 0$. First, we generalize the existing polymatroidal outer bound on the capacity region of the centralized problem to the distributed case. Next, building upon the existing centralized composite coding scheme, we propose three distributed composite coding schemes and derive the corresponding inner bounds on the capacity region. We present a number of interesting numerical examples, which highlight the subtleties and challenges of dealing with the distributed index coding, even for very small problem sizes of $n=3$ and $n=4$.
\end{abstract}

\section{Introduction}\label{sec:intro}

The index coding problem is a canonical problem in network information theory and has been studied over the past two decades using tools from various disciplines such as combinatorics, algebra, and information theory \cite{Birk--Kol2006,Effros--El-Rouayheb--Langberg2013}.
In the traditional index coding problem, it is assumed that one server has all messages requested by the receivers.
However, in many practical circumstances this assumption might not be true and the messages might be distributed over multiple servers.
We refer to this more general version as the {\em distributed index coding problem} compared to the traditional {\em centralized} index coding problem. 
The distributed index coding problem was first studied by Ong, Ho, and Lim \cite{Ong--Ho--Lim2014}, where lower and upper bounds on the optimal codelength were derived in the special case in which each message is only known to one receiver and each receiver only knows one message a priori and it is shown that the bounds match if no two servers have any messages in common.

In this paper, we study the distributed index coding problem in its general form, which to the best of our knowledge has not been investigated before.
First, we generalize the polymatroidal outer bound on the capacity region of the centralized problem \cite{Arbabjolfaei--Bandemer--Kim--Sasoglu--Wang2013} to the distributed case.
Next, building upon the existing centralized composite coding scheme \cite{Arbabjolfaei--Bandemer--Kim--Sasoglu--Wang2013}, we propose three distributed composite coding schemes, derive the corresponding inner bounds on the capacity region and show their use via examples. Although the polymatroidal outer bound is tight for all centralized index coding problems with up to $n=5$ messages \cite{Arbabjolfaei--Bandemer--Kim--Sasoglu--Wang2013}, there exist instances of the distributed problem with $n=3$ messages for which the polymatroidal outer bound is not tight. 
Nevertheless, using customized Shannon-type inequalities, we show that the proposed distributed composite coding scheme  achieves the capacity region for all instances of the problem with $n=3$ messages.


In this paper, $[n]$ denotes the set $\{1, 2, \ldots, n\}$ and the set of all nonempty subsets of $[n]$ is 
$N \doteq \{J \subseteq [n] \suchthat J \not = \emptyset\}.$

\section{System Model and Problem Setup}\label{sec:model}

\begin{figure}[b]
\begin{center}
\small
\psfrag{1}[cb]{1}
\psfrag{2}[rc]{2}
\psfrag{3}[lc]{3}
\psfrag{4}{4}
\includegraphics[scale=0.35]{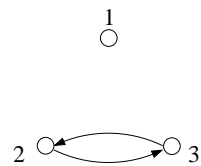}
\end{center}
\caption{A 3-node side information graph representing the index coding problem with $A_1 = \emptyset, A_2 = \{3\}$, and $A_3 = \{2\}$. 
}
\label{fig:3-message}
\end{figure}

Consider the following index coding problem. There are $n$ messages in the system, $x_1, x_2, \cdots, x_n$, where $x_j \in \{0,1\}^{t_j}$ for $j \in [n]$ and some $t_j$. There are $n$ receivers, where receiver $j$ wants to obtain message $x_j$ and knows a subset of the messages a priori, denoted by $x(A_j)$ for some $A_j \subseteq [n]\setminus \{j\}$. For simplicity of notation throughout the paper and where the context is clear,  we will refer to $j$ as the wanted message and to $A_j$ as the side information of receiver $j$, respectively.
Any instance of this problem can be specified by a side information graph $G$ with $n$ nodes, in which a directed edge $i \to j$ represents that receiver $j$ has message $i$ as side information ($i \in A_j$).
For instance, Fig.~\ref{fig:3-message} shows the directed graph representing the index coding problem with $A_1 = \emptyset$, $A_2 = \{3\}$, and $A_3=\{2\}$.

The main difference in the system model compared to traditional (centralized) index coding problem is in the server setup. Instead of a single server which contains all messages,  there are $2^n-1$ servers. For each $J \in N$, there is a server that contains all messages $j \in J$ and the capacity of the broadcast link connecting server $J$ to all receivers is denoted by $C_J$. Hence, we assume that there are $2^n-1$ ideal bit pipes to the receivers with arbitrary link capacities. This is a fairly general model that allows for all possible message availabilities on different servers. If $C_J = 1$ only for $J= [n]$ and is zero otherwise, we recover the centralized index coding problem. A special normalized symmetric case is where $C_J = 1$ for all $J \in N$. Server $J$ sends sequence $y_J \in \{0,1\}^{u_J}$ for some $u_J$ to all receivers which is a function of the messages at that server.


\begin{figure}[t!]
\begin{center}
\small
\psfrag{x1}[b]{$x_1$}
\psfrag{x2}[b]{$x_2$}
\psfrag{x3}[b]{$x_3$}
\psfrag{x12}[b]{$x_1, x_2$}
\psfrag{x13}[b]{$x_1, x_3$}
\psfrag{x23}[b]{$x_2, x_3$}
\psfrag{x123}[b]{$x_1, x_2, x_3$}
\psfrag{y1}[b]{$y_{\{1\}}$}
\psfrag{y2}[b]{$y_{\{2\}}$}
\psfrag{y3}[b]{$y_{\{3\}}$}
\psfrag{y12}[b]{$y_{\{1,2\}}$}
\psfrag{y13}[b]{$y_{\{1,3\}}$}
\psfrag{y23}[b]{$y_{\{2,3\}}$}
\psfrag{y123}[b]{$y_{\{1,2,3\}}$}
\psfrag{e1}[c]{Server $\{1\}$}
\psfrag{e2}[c]{Server $\{2\}$}
\psfrag{e3}[c]{Server $\{3\}$}
\psfrag{e12}[c]{Server $\{1,2\}$}
\psfrag{e13}[c]{Server $\{1,3\}$}
\psfrag{e23}[c]{Server $\{2,3\}$}
\psfrag{e123}[c]{Server $\{1,2,3\}$}
\psfrag{d1}[c]{Decoder $1$}
\psfrag{d2}[c]{Decoder $2$}
\psfrag{d3}[c]{Decoder $3$}
\psfrag{xh1}[b]{$\xh_1$}
\psfrag{xh2}[b]{$\xh_2$}
\psfrag{xh3}[b]{$\xh_3$}
\psfrag{a1}[b]{$x({A_1})$}
\psfrag{a2}[b]{$x({A_2})$}
\psfrag{a3}[b]{$x({A_3})$}
\includegraphics[scale=0.35]{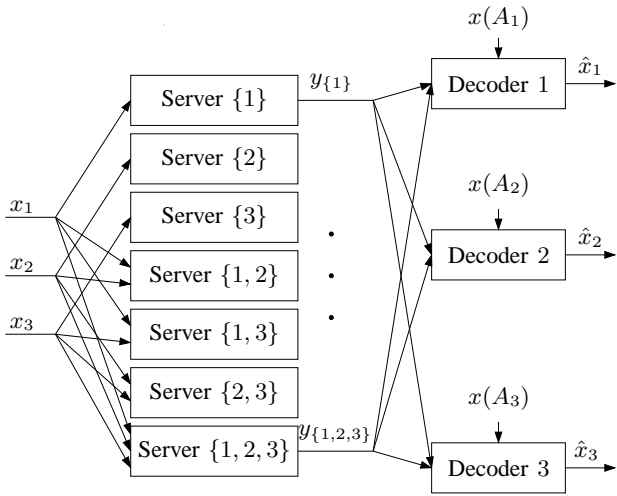}
\end{center}
\caption{The distributed index coding problem with $n=3$.}
\label{fig:index_coding}
\end{figure}

Based on the side information $A_j$ and the received bits $y_J \in \{0,1\}^{u_J}$ from all servers, receiver $j$ finds the estimate $\hat{x}_j$ of the message $x_j$ (see Fig.~\ref{fig:index_coding}). We say that rate-capacity tuple $(\mathbf{R}, \mathbf{C}) = ((R_j, j \in [n]),(C_J: J \in N))$ is achievable if there exists $r$ such that:	
\begin{align}
R_j \le \frac{t_j}{r},  \quad C_J \ge \frac{u_J}{r}, \quad \forall j \in [n], \quad \forall J \in N.
\end{align}
For a given $\mathbf{C}$, the capacity region $\Cr$ of this index coding problem is the closure of the set of achievable rate tuples $\mathbf{R} = (R_1, \cdots, R_n)$. 


Throughout the paper, we will compactly represent the distributed index coding problem (for a given $\mathbf{C}$) as sets of $(j|i \in A_j)$. For example, for $A_1 = \emptyset$, $A_2 = \{3\}$, and $A_3=\{2\}$, we write
$(1); (2|3); (3|2)$.

\section{Outer Bounds}\label{sec:outer:bounds}


In this section, we generalize the polymatroidal outer bound for the centralized index coding problem (see, for example, {\cite[Theorem~1]{Arbabjolfaei--Bandemer--Kim--Sasoglu--Wang2013}})  to the distributed case.

\begin{theorem}
\label{thm:polymatroidal}
Let $B_j$ be the set of interfering messages at receiver $j$, i.e., $B_j = [n] \setminus (A_j \cup \{j\})$.
If $(\Rv, \Cv)$ is achievable, then for every $T \in N$,
\begin{align}
R_j \leq f_T(B_j \cup \{j\}) - f_T(B_j), \quad j \in T, 
\end{align}
for some $f_T(S)$, $S \subseteq T$, such that
\begin{enumerate}
\item $f_T(\emptyset) = 0$,
\item $f_T(T) = \sum_{J: J \cap T \not = \emptyset} C_J$,
\item $f_T(A) \leq f_T(B)$ for all $A \subseteq B \subseteq T$, and
\item $f_T(A \cup B) + f_T(A \cap B) \leq f_T(A) + f_T(B), \forall A,B \subseteq T$. 
\end{enumerate} 
\end{theorem}

A relaxed version of this bound, which is a generalized version of the maximal acyclic induced subgraph (MAIS) bound, is handy and also useful for some problems.
\begin{corollary}
\label{coro:mais}
If $(\Rv,\Cv)$ is achievable for an index coding problem represented by the directed graph $G$, then for every $T \in N$ it
must satisfy
\begin{align}
\sum_{j \in S} R_j \leq \sum_{J: J \cap T \not = \emptyset} C_J,
\end{align}
for all $S \subseteq T$ for which the subgraph of $G$ induced by $S$ does not contain a directed cycle.
\end{corollary}

\begin{example}
\label{ex:3-node}
Consider the  index coding problem shown in Fig.~\ref{fig:3-message}.
The following is the generalized MAIS outer bound for this problem (inactive inequalities are not shown).
\begin{align*}
R_1 &\leq C_{\{1\}} + C_{\{1,2\}} + C_{\{1,3\}} + C_{\{1,2,3\}}, \\
R_2 &\leq C_{\{2\}} + C_{\{1,2\}} + C_{\{2,3\}} + C_{\{1,2,3\}}, \\
R_3 &\leq C_{\{3\}} + C_{\{1,3\}} + C_{\{2,3\}} + C_{\{1,2,3\}}, \\
R_1 + R_2 &\leq  C_{\{1\}} + C_{\{2\}} + C_{\{1,2\}} + C_{\{1,3\}} + C_{\{2,3\}} + C_{\{1,2,3\}}, \\
R_1 + R_3 &\leq  C_{\{1\}} + C_{\{3\}} + C_{\{1,2\}} + C_{\{1,3\}} + C_{\{2,3\}} + C_{\{1,2,3\}}.
\end{align*}
\end{example}

\section{Inner Bounds}\label{sec:inner:bounds}
\def\R{\mathcal R}
\newcommand{\ckj}{C_{K,J}}
In this section, we present a series of inner bounds on the capacity region of the distributed index coding problem, all of which are built upon the method of composite index coding, first introduced in \cite{Arbabjolfaei--Bandemer--Kim--Sasoglu--Wang2013}. The first inner bound is a simple extension of \cite{Arbabjolfaei--Bandemer--Kim--Sasoglu--Wang2013}, in which we solve the composite coding problem separately for each server. We show the shortcoming of this method through an example and then present a new composite coding scheme that solves the problem collectively across all servers. The limitation of this scheme, which is assigning the same decoding sets across all servers, is demonstrated through an example. This leads us to the last inner bounding scheme as a generalization of the two previous methods, where we allow a general grouping of servers for solving  distributed index coding.   

In this section, we assume that for each non-empty subset $K \subseteq J$ at server $J \in N$, there is a virtual encoder at that server with associated composite coding rate $\ckj$.

\subsection{Combining Centralized Composite Coding Inner Bounds}\label{sec:separate}
The idea is to separately apply the composite coding inner bound to each server and appropriately combine the achievable rates. Fix server $J$ and for $j \notin J$, set $R_j = 0$. For every $j \in J$, let $D_{j,J}$ be the index of messages that receiver $j$ wishes to decode from server $J$ and $A_{j,J} = A_j \cap J$. The rates of the composite messages belong to the polymatroidal rate region $\R(D_{j,J}|A_{j,J})$ defined by            
\begin{align}
\sum_{j \in \textcolor{black}{L_J}} R_{j,J} <\sum_{K\subseteq D_{j,J}\cup A_{j,J}:\,K\cap L_J \neq \emptyset} \ckj,
\end{align}
for all $L_J \subseteq D_{j,J}\setminus A_{j,J}$. Then the achievable rate region for server $J$ is given by
\begin{align}
\mathbf{R}_J \in \bigcap_{j \in J}\quad\bigcup_{D_{j,J}\subseteq J:j \in D_{j,J}} \R(D_{j,J}|A_{j,J}),
\end{align}
with the following constraint on composite rates of server $J$
           \begin{align}\sum_{K:K \not \subseteq A_j} \ckj \le C_{J}, \quad \forall j \in J.\end{align}
            
                       After separately finding the composite coding  rate regions for all servers, we add the corresponding constraints. Note that we need to write all possible rate equalities and inequalities, including the inactive ones. Table \ref{tabel:individual3} shows the application of this scheme to the problem $(1|3);(2|1);(3|2)$ when $C_J = 1$ for all $J \in N$. We note that the sum rate $R_1+R_2+R_3$ is limited to 7.5. However, the outer bound of Theorem \ref{thm:polymatroidal} gives a sum rate of $R_1+R_2+R_3 <9$. In the next subsection, we will see that this outer bound can be achieved 
            by treating servers collectively.
            
 
\setlength{\tabcolsep}{4pt}
\definecolor{light-gray}{gray}{0.9}
\hspace{10mm}
\rowcolors{1}{light-gray}{white}
\begin{table*}
\begin{center}
\caption{Rate region for problem $(1|3);(2|1);(3|2)$ with combining centralized composite coding.} 
\footnotesize
\begin{tabular}{|c| c| c| c| c| c| c| c|}
\myhline
\opertype{\textbf{Server Index $J$}}&$R_1$&$R_2$&$R_3$&$R_1+R_2$&$R_1+R_3$&$R_2+R_3$&$R_1+R_2+R_3$\\
\myhline
\{1\}&$<1$&$=0$&$=0$&$<1$&$<1$&$=0$&$<1$\\
\myhline
\{2\}&$=0$&$<1$&$=0$&$<1$&$=0$&$<1$&$<1$\\
\myhline
\{3\}&$=0$&$=0$&$<1$&$=0$&$<1$&$<1$&$<1$\\
\myhline
\{1,2\}&$<1$&$<1$&$=0$&$<1$&$<1$&$<1$&$<1$\\
\myhline
\{1,3\}&$<1$&$=0$&$<1$&$<1$&$<1$&$<1$&$<1$\\
\myhline
\{2,3\}&$=0$&$<1$&$<1$&$<1$&$<1$&$<1$&$<1$\\
\myhline
\{1,2,3\}&$<1$&$<1$&$<1$&$<1$&$<1$&$<1$&$<1.5$\\
\myhline
\textbf{Sum rates}&$<4$&$<4$&$<4$&$<6$&$<6$&$<6$&$<7.5$\\
\myhline
\end{tabular}
\label{tabel:individual3}
\end{center}
\vspace{-0.5cm}
\end{table*}
\normalsize
\vspace{-4mm}
\subsection{Distributed Composite Coding with Same Decoding Sets}\label{sec:allservers}
Here, we fix a decoding set $D_j$ for receiver $j$ across all servers and solve a single distributed coding problem. Intuitively, the advantage of this scheme is that receivers can collectively use servers to decode messages, even though some of those messages do not exist on some servers. 

Let $D_j$ be the index of messages receiver $j$ decodes.  The rates of the composite messages belong to the polymatroidal rate region $\R(D_{j}|A_{j})$ defined by      
\begin{align}
\sum_{j \in \textcolor{black}{L}} R_{j,[n]} <\sum_{K\subseteq D_j\cup A_j:K\cap L \neq \emptyset}\quad \sum_{J: K \subseteq J} \ckj,
\end{align}
for all $L \subseteq D_j\setminus A_j$. The second sum on the RHS is because all servers that contain $K$ contribute to the composite rate. Similar as before, we have
\begin{align}
\mathbf{R}_{[n]} \in \bigcap_{j\in[n]}\quad\bigcup_{D_j\subseteq [n]:j \in D_j} \R(D_j|A_j),
\end{align}
and the constraints on the composite rates are 
            \begin{align}\sum_{K:K \not \subseteq A_j} \ckj \le C_{J}, \quad \forall j \in [n], \forall J \in N.\end{align}
            
Table \ref{tabel:all3} shows the composite coding inner bound along with an optimal decoding set for all non-isomorphic cases with $n=3$ and when $C_J = 1$ for all $J \in N$. It can be seen that for the problem  $(1|3);(2|1);(3|2)$ studied in Table \ref{tabel:individual3}, the inner bound matches the MAIS outer bound, hence establishing the capacity region. 
In fact, for all non-isomorphic cases shown in Table \ref{tabel:all3}, the composite coding inner bound matches the MAIS outer bound, except for the cases that are in the same group as the problem $(1);(2|3);(3|2)$.
For these instances, the 
inequality 
\begin{align}
\label{eq:outerineq}
R_1 + R_2 + R_3 \leq 9
\end{align}
is not deducible by Theorem \ref{thm:polymatroidal}.
However, using customized Shannon-type inequalities, we show that this inequality is also an outer bound inequality (see Appendix \ref{app:outerineq}).
Therefore, for all non-isomorphic index coding problems with $n=3$, the rates shown in Table \ref{tabel:all3} are indeed capacity regions.
Although the capacity regions are only shown for the case of $C_J = 1$, $J \in N$, it can be easily verified that the capacity regions of all problems with $n=3$ are achieved by the composite coding scheme for arbitrary values of $C_J$, $J \in N$.
Furthermore, in Table \ref{tabel:all3} we can see that the capacity regions of the two problems $(1);(2|3);(3|2)$ and $(1|3);(2|3);(3|2)$ are not the same.
This proves the important point that, as opposed to the structural property of the centralized index coding \cite{Gadouleau--Riis2011}, \cite{Tahmasbi--Shahrasbi--Gohari2014b}, \cite{Arbabjolfaei--Kim2015a}, removing an edge that does not belong to any directed cycle may change the capacity region of the distributed index coding problem.


\setlength{\tabcolsep}{2pt}
\definecolor{light-gray}{gray}{0.9}
\renewcommand{\opertype}[1]{\begin{minipage}{48mm}\centering\vspace{1mm} #1\vspace{1mm}\end{minipage}}
\hspace{10mm}
\rowcolors{1}{light-gray}{white}
\footnotesize
\begin{table*}
\begin{center}
\caption{Capacity region for all non-isomorphic problems of size $n=3$.}
\begin{tabular}{|c| c| c| c| c| c| c| c| c|}

\myhline
\opertype{\textbf{IC Problem}}&$R_1$&$R_2$&$R_3$&$R_1+R_2$&$R_1+R_3$&$R_2+R_3$&$R_1+R_2+R_3$&\opertype{\textbf{An Optimal Decoding Set}}\\
\myhline
\opertype{$(1);(2);(3)\quad\,\,\, (1|2);(2);(3)$\\$(1|2,3);(2);(3)\,\, (1);(2|3);(3|1)$\\$(1);(2|1);(3|1)\quad\,\,\,(1);(2|1);(3|1,2)$}&$\leq4$&$\leq4$&$\leq4$&$\leq6$&$\leq6$&$\leq6$&$\leq7$&\opertype{$D_j = [n]\setminus A_j$}\\
								\myhline
						\opertype{$(1|3);(2|1);(3|2)$}&$\leq4$&$\leq4$&$\leq4$&$\leq6$&$\leq6$&$\leq6$&$\leq9$&\opertype{$D_j = [n]\setminus A_j$}\\
		\myhline
		
		\opertype{$(1);(2|3);(3|2)\quad(1);(2|1,3);(3|2)$\\$(1);(2|1,3);(3|1,2)$}&$\leq4$&$\leq4$&$\leq4$&$\leq6$&$\leq6$&$\leq8$&$\leq9$&\opertype{If $A_j = \emptyset$ then $D_j = \{j\}$, otherwise $D_j = [n]\setminus A_j$}\\
		\myhline
				\opertype{$(1|3);(2|3);(3|2)\quad(1|3);(2|1,3);(3|2)$\\$(1|2,3);(2|3);(3|2)$}&$\leq4$&$\leq4$&$\leq4$&$\leq6$&$\leq6$&$\leq8$&$\leq10$&\opertype{$D_j = [n]\setminus A_j$}\\
		\myhline
								\opertype{$(1|3);(2|3);(3|1,2)$\\$(1|2,3);(2|3);(3|1,2)$}&$\leq4$&$\leq4$&$\leq4$&$\leq6$&$\leq8$&$\leq8$&$\leq10$&\opertype{$D_j = [n]\setminus A_j$}\\
		\myhline
\opertype{$(1|2,3);(2|1,3);(3|1,2)$}&$\leq4$&$\leq4$&$\leq4$&$\leq8$&$\leq8$&$\leq8$&$\leq12$&\opertype{$D_j = [n]\setminus A_j$}\\
		\myhline
\end{tabular}
\label{tabel:all3}
\end{center}
\vspace{-0.5cm}	
\end{table*}
\normalsize \hspace{-10mm}To show the limitation of this composite coding scheme, consider the problem $(1|4); (2|3,4); (3|1,2);(4|2,3)$. We find $R_1+R_2+R_3+R_4<23$ to be the best achievable sum rate under this scheme with the full inner bound characterized as
\small
   \begin{align*}
R_1&<8, \quad R_1+R_2<12\\
R_2&<8, \quad R_1+R_3<12\\
 R_3&<8, \quad R_1+R_4<12\\
R_4&<8, \quad R_3+R_4<12\\
R_1+R_2+R_3&<18, \,\,R_1+R_2+R_3+R_4<23
   \end{align*}  
   \normalsize
which is determined by setting $D_1 = \{1\}$, $D_2 = \{1,2\}$, $D_3 = \{3,4\}$ and $D_4 = \{1,4\}$ as in its centralized index coding counterpart studied in \cite{Arbabjolfaei--Bandemer--Kim--Sasoglu--Wang2013}. However, the outer bound of Theorem \ref{thm:polymatroidal} gives $R_1+R_2+R_3+R_4\le 24$. In the next subsection, we will show that indeed this sum rate can be achieved and the capacity region can be established 
   by appropriately grouping servers and assigning suitable different decoding sets to each group.


\subsection{Distributed Composite Coding with Different Decoding Sets}

Intuitively, this scheme can be thought of as a generalization of the previous two schemes where index coding was solved by treating servers either  separately or collectively. In this new scheme, we jointly optimize grouping of servers and decoding sets across server groupings. 
 
 Consider a subset of servers 
 $P \subseteq N$ and let $J' = \cup_{J\in P} J$ be the union of all messages held by servers in $P$. For $j \notin J'$, we set $R_j = 0$. For every $j \in J'$, let $D_{j,P}$ be the index of messages that receiver $j$ wishes to decode from servers in $P$ and $A_{j,P} = \cup_{J\in P} (A_j \cap J)$.  The rates of the composite messages belong to the polymatroidal region $\R(D_{j,P}|A_{j,P})$:
\begin{align*}
\sum_{j \in \textcolor{black}{L_P}} R_{j,P} <\sum_{K\subseteq D_{j,P}\cup A_{j,P}:\,K\cap L_J \neq \emptyset} \quad \sum_{J: J \in P, K \subseteq J} \ckj,
\end{align*}
for all $L_P \subseteq D_{j,P}\setminus A_{j,P}$.  Then the achievable rate region for servers in $P$ is given by
\begin{align}
\mathbf{R}_P \in \bigcap_{j\in J'}\quad\bigcup_{D_{j,P}\subseteq J':j \in D_{j,P}} \R(D_{j,P}|A_{j,P}),
\end{align}
with the following constraint on composite rates 
            \begin{align}\sum_{K:K \not \subseteq A_j} \ckj \le C_{J}, \quad \forall j \in J', \forall J \in P.\end{align}
                After finding the composite coding  rate regions for all server groupings, we appropriately add the corresponding constraints. Note that in this scheme we need to write all possible rate equalities and inequalities, including the inactive ones. Also, note that the previous two composite coding schemes are special cases of this method with $2^n-1$ server groups of the form $P = J$ in Section \ref{sec:separate} and a single server group of $P = N$ in Section \ref{sec:allservers}, respectively.
                
The scheme is best shown via the example of Table \ref{tabel:4optimal} where we revisit the problem $(1|4); (2|3,4); (3|1,2);(4|2,3)$. Each row represents a server grouping and the corresponding decoding sets are shown in the last column. Note for example, how the decoding set for receiver $1$ varies across different server groupings. The obtained inner bound matches the MAIS outer bound, hence establishing the capacity region. 

\setlength{\tabcolsep}{5pt}
\definecolor{light-gray}{gray}{0.9}
\renewcommand{\opertype}[1]{\begin{minipage}{29mm}\centering\vspace{1mm} #1\vspace{1mm}\end{minipage}}
\rowcolors{1}{light-gray}{white}
 \footnotesize
\begin{table*}
\begin{center}
	\caption{Capacity region for the problem $(1|4); (2|3,4); (3|1,2);(4|2,3)$ is achieved via server groupings.}
\begin{tabular}{|c| c| c| c| c| c| c| c| c| c|}
\myhline
\opertype{\textbf{Server Groupings}}&$R_1$&$R_2$&$R_3$&$R_4$&$R_1+R_2$&$R_1+R_3$&$R_1+R_4$&$R_3+R_4$&\opertype{\textbf{Relevant Optimal Decoding Sets}}\\
\myhline
P = \{\{1\}\}&$\leq1$&$=0$&$=0$&$=0$&$\leq1$&$\leq1$&$\leq1$&$=0$&$D_1 = \{1\}$\\
\myhline
P = \{\{2\}\}&$=0$&$\leq1$&$=0$&$=0$&$\leq1$&$=0$&$=0$&$=0$&$D_2 = \{2\}$\\
\myhline
P = \{\{3\}\}&$=0$&$=0$&$\leq1$&$=0$&$=0$&$\leq1$&$=0$&$\leq1$& $D_3 = \{3\}$\\
\myhline
P = \{\{4\}\}&$=0$&$=0$&$=0$&$\leq1$&$=0$&$=0$&$\leq1$&$\leq1$& $D_4 = \{4\}$\\
\myhline
P = \{\{1,2\}\}&$\leq1$&$\leq1$&$=0$&$=0$&$\leq1$&$\leq1$&$\leq1$&$=0$&$D_1 = \{1,2\}, D_2 = \{1,2\}$\\
\myhline
$P = \{\{1,3\}, \{3,4\}\}$ 	&$\leq1$&$=0$&$\leq2$&$\leq1$&$\leq1$&$\leq2$&$\leq2$&$\leq2$&$D_1 = \{1,3\}, D_3 = \{3,4\}, D_ 4 =\{ 1,4\}$\\
\myhline
$P = \{\{2,3\}, \{2,4\}\}$  	&$=0$&$\leq2$&$\leq1$&$\leq1$&$\leq2$&$\leq1$&$\leq1$&$\leq2$&$D_2 = \{2\}, D_ 3 =\{ 3,4\}, D_ 4 =\{ 4\}$\\
\myhline
\opertype{$P = \{\{1,4\}, \{1,2,3,4\},$\\$\{1,2,3\}, \{1,2,4\},$\\$\{1,3,4\}, \{2,3,4\}\}$}	&$\leq5$&$\leq4$&$\leq4$&$\leq5$&$\leq6$&$\leq6$&$\leq6$&$\leq6$&\opertype{$D_1 = \{1\},$\\ $D_2 = \{1,2\},$\\ $D_3 = \{3,4\},$\\ $D_ 4 =\{ 1,4\}$}\\
		\myhline
\textbf{Sum rates}&$\leq8$&$\leq8$&$\leq8$&$\leq8$&$\leq12$&$\leq12$&$\leq12$&$\leq12$&\\
\myhline
	\end{tabular}\label{tabel:4optimal}
	\end{center}
	\vspace{-0.5cm}
	\end{table*}
\normalsize

\section{Conclusion}\label{sec:conclusion}
\vspace{-3mm}
We studied a more general class of index coding problems than those considered in the literature, where for a given problem of size $n$ we allowed all possible distributed servers with arbitrary broadcast link capacities. We derived a generalized polymatroidal outer bound on the capacity region. We also proposed a new general composite coding scheme and showed the necessity of using this general scheme to achieve the distributed index coding capacity region for a problem of size $n=4$. We showed that even for small distributed index coding problems of size $n=3$, there exist instances where the generalized polymatroidal outer bound is not tight and customized Shannon-type inequalities are required to establish the capacity region. Moreover, 
we showed instances of the distributed index coding problems of size $n=3$, whose graph structural properties are fundamentally different from their centralized index coding counterparts.

There are many interesting directions for future research work on this challenging problem. One direction is to derive tighter outer bounds. The other general direction is to study the graph structural properties of the problem and other graph-based inner and outer bounds.
\vspace{-2mm}
\appendices
\section{Proof of Theorem \ref{thm:polymatroidal}}
\label{app:polymatroidal}

Consider a $(t_1, \ldots, t_n, r)$ index code for the problem $(j|A_j)$, $j \in [n]$.
Let $X_j$ be the uniform random variable over $\{0,1\}^{t_j}$ corresponding to message $j \in [n]$ and $Y_J$ be the uniform random variable over $\{0,1\}^{u_J}$ corresponding to the output of server $J \in N$.
For each $T \in N$ we set $x_j = \emptyset$, $j \not \in T$.
Therefore, $Y_J = \emptyset$, if $J \cap T = \emptyset$.
Let $Y \doteq (Y_J, J \cap T \not = \emptyset)$.
For $j \in T$
\begin{align*}
t_j = H(X_j) &= I(X_j;Y | X(A_j \cap T)) \\
& = H(Y | X(A_j \cap T)) - H(Y | X_j,X(A_j \cap T)).
\end{align*}
Define
\begin{align}
\label{eq:f-def2}
f_T(S) = \sum_{J: J \cap T \not = \emptyset} C_J ~~~ \frac{H(Y  | X(S^c \cap T))}{H(Y)} .
\end{align}
Then the set function $f_T$ satisfies
$f_T(\emptyset) = 0$, and
$f_T(T) = \sum_{J: J \cap T \not = \emptyset} C_J$.
Let $A \subseteq B \subseteq T$ ($B^c \subseteq A^c$), then $f_T(A) \leq f_T(B)$ since conditioning reduces entropy.
We show that  the set function $f_T$ is submodular.
Consider $A, B \subseteq T$, then
\begin{align}
&f_T(A \cup B) - f_T(A) = \frac{\sum_{J: J \cap T \not = \emptyset} C_J}{H(Y)} \times \nonumber \\
\label{eq:left-hs}
& ( H(Y| X(A^c \cap B^c \cap T)) -  H(Y|X(A^c \cap T)) ),
\end{align}
\begin{align}
\label{eq:right-hs}
&f_T(B) - f_T(A \cap B) = \frac{\sum_{J: J \cap T \not = \emptyset} C_J}{H(Y)} \times \nonumber \\
& ( H(Y| X(B^c \cap T)) -  H(Y|X((A^c \cup B^c) \cap T)) ).
\end{align}
We have
\begin{align}
&H(Y| X(A^c \cap B^c \cap T)) -  H(Y|X(A^c \cap T)) \nonumber \\
&= I(Y; X(A^c \cap B \cap T) | X(A^c \cap B^c \cap T)) \nonumber \\
\label{eq:1}
&= H(X(A^c \cap B \cap T)) \nonumber \\
&- H(X(A^c \cap B \cap T) | Y, X(A^c \cap B^c \cap T)),
\end{align}
and similarly
\begin{align}
&H(Y| X(B^c \cap T)) -  H(Y|X((A^c \cup B^c) \cap T)) \nonumber \\
\label{eq:2}
&= H(X(A^c \cap B \cap T)) \nonumber \\
&- H(X(A^c \cap B \cap T)| Y, X(B^c \cap T)).
\end{align}
Since conditioning reduces entropy, \eqref{eq:left-hs}-\eqref{eq:2} imply
\begin{align*}
f_T(A \cup B) - f_T(A) \leq f_T(B) - f_T(A \cap B), 
\end{align*}
which completes the proof of the submodularity of the set function $f_T$ defined in \eqref{eq:f-def2}.
Let $(\Rv, \Cv)$ be achievable. Then for every $T \in N$ and every $j \in T$ we have
\begin{align*}
R_j \leq \frac{t_j}{r} &= \left(f_T(B_j \cup \{j\}) - f_T(B_j)\right) \frac{H(Y)}{r \sum_{J: J \cap T \not = \emptyset} C_J} \\
&\leq \left(f_T(B_j \cup \{j\}) - f_T(B_j)\right) \frac{\sum_{J: J \cap T \not = \emptyset} u_J}{r \sum_{J: J \cap T \not = \emptyset} C_J} \\
&\leq f_T(B_j \cup \{j\}) - f_T(B_j).
\end{align*}

\section{Proof of inequality \eqref{eq:outerineq}}
\label{app:outerineq}

Consider a $(t_1, t_2, t_3, r)$ index code for the problem $(1);(2|3);(3|2)$.
Let $X_j$ be the uniform random variable over $\{0,1\}^{t_j}$ corresponding to message $j \in \{1,2,3\}$ and $Y_J$ be the uniform random variable over $\{0,1\}^{u_J}$ corresponding to the output of server 
$J \in N_3 = \{J \subseteq \{1,2,3\}, J \not = \emptyset\}$.
Since the output of every server is a function of the messages available at that server, 
\begin{align}
\label{eq:111}
H(Y_J | X_i, i \in J) &= 0, \quad J \in N_3,
\end{align}
and due to the exact recovery condition at each receiver
\begin{align}
\label{eq:222}
H(X_1|Y_J, J \in N_3) &= 0, \\
\label{eq:333}
H(X_2|(Y_J, J \in N_3), X_3) &= 0, \\
\label{eq:444}
H(X_3|(Y_J, J \in N_3) ,X_2) &= 0.
\end{align}
The independence of the messages and \eqref{eq:111}-\eqref{eq:444} leads to
\begin{align}
\label{eq:eq1}
&H(X_2 \cond X_1) = H(X_2 \cond X_1, X_3) \\
&= H(X_2 \cond  X_1, X_3, Y_{\{1\}}, Y_{\{3\}}, Y_{\{1,3\}}) \nonumber \\
&= H(X_2, Y_{\{2\}}, Y_{\{1,2\}}, Y_{\{2,3\}}, Y_{\{1,2,3\}} \cond  X_1, X_3, Y_{\{1\}}, Y_{\{3\}}, Y_{\{1,3\}}) \nonumber \\
&= H(Y_{\{2\}}, Y_{\{1,2\}}, Y_{\{2,3\}}, Y_{\{1,2,3\}} \cond  X_1, X_3, Y_{\{1\}}, Y_{\{3\}}, Y_{\{1,3\}}) \nonumber \\
&\leq H(Y_{\{2\}}, Y_{\{1,2\}}, Y_{\{2,3\}}, Y_{\{1,2,3\}} \cond  X_1). \nonumber
\end{align}
Next consider
\begin{align}
\label{eq:eq2}
&I(X_2;Y_{\{2\}}, Y_{\{1,2\}}, Y_{\{2,3\}}, Y_{\{1,2,3\}} \cond X_1) \\
&= H(X_2 \cond X_1) - H(X_2 \cond Y_{\{2\}}, Y_{\{1,2\}}, Y_{\{2,3\}}, Y_{\{1,2,3\}}, X_1) \nonumber \\
&= H(Y_{\{2\}}, Y_{\{1,2\}}, Y_{\{2,3\}}, Y_{\{1,2,3\}} \cond X_1) \nonumber\\&\quad - H(Y_{\{2,3\}}, Y_{\{1,2,3\}} \cond X_1,X_2).
\end{align}
Comparing \eqref{eq:eq1} and \eqref{eq:eq2} we get
\begin{align}
\label{eq:eq3}
&H(X_2 \cond Y_{\{2\}}, Y_{\{1,2\}}, Y_{\{2,3\}}, Y_{\{1,2,3\}}, X_1) \nonumber \\
&\leq H(Y_{\{2,3\}}, Y_{\{1,2,3\}} \cond X_1,X_2).
\end{align}
We have
\begin{align}
&H(X_1, X_2, X_3 \cond Y_J, J \in N_3) 
= H(X_2 \cond (Y_J, J \in N_3), X_1) \nonumber \\
\label{eq:eq4}
&\leq H(Y_{\{2,3\}}, Y_{\{1,2,3\}} \cond X_1,X_2) \\
\label{eq:eq5}
&\leq H(Y_{\{2,3\}}, Y_{\{1,2,3\}}),
\end{align}
where \eqref{eq:eq4} follows from \eqref{eq:eq3}.
\begin{align}
& t_1 + t_2 + t_3
= H(X_1,X_2,X_3) \nonumber \\
&= H(X_1,X_2,X_3) + H(Y_J, J \in N_3 \cond X_1,X_2,X_3) \nonumber \\
&= H(Y_J, J \in N_3) + H(X_1, X_2, X_3 \cond Y_J, J \in N_3) \nonumber \\
\label{eq:eq6}
&\leq H(Y_J, J \in N_3) + H(Y_{\{2,3\}}, Y_{\{1,2,3\}}) \\
&\leq \sum_{J \in N_3} u_J + u_{\{2,3\}} + u_{\{1,2,3\}} \nonumber \\
&\leq r \left( \sum_{J \in N_3} C_J + C_{\{2,3\}} + C_{\{1,2,3\}} \right)
\label{eq:eq7} = 9r,
\end{align}
where \eqref{eq:eq6} follows from \eqref{eq:eq5} and \eqref{eq:eq7} follows from the assumption of $C_J = 1$, $J \in N_3$.
Therefore,
$
R_1 + R_2 + R_3 \leq 9.
$


%
%
%
%
%
%
%
%
%
%
%


\newcommand{\noopsort}[1]{}



\end{document}